\documentclass[12pt]{article}
\usepackage{amsmath}
\usepackage{graphicx}
\usepackage{natbib}
\usepackage{url} 
\usepackage[title]{appendix}
\usepackage{float}
\usepackage[colorinlistoftodos]{todonotes}
\usepackage[colorlinks=true, allcolors=blue]{hyperref}
\usepackage{listings}
\usepackage{url}
\usepackage{bm}
\usepackage{setspace}
\usepackage{subfig}
\usepackage{algorithm}
\usepackage{algpseudocode}
\usepackage{amssymb}
\usepackage{csquotes}
\usepackage{multirow}
\usepackage{cancel}

\newcommand{\numOfDG}{K} 
\newcommand{\colOfDG}{\bm{\Lambda}} 
\newcommand{\indOfDG}{k}
\newcommand{\eleOfDG}{\lambda}

\newcommand{\numOfThr}{H}
\newcommand{\colOfThr}{\bm{\mathcal{H}}}
\newcommand{\indOfThr}{h}
\newcommand{\eleOfThr}{\eta}
\newcommand{\rvOfThr}{\eta}

\newcommand{\numOfReg}{R}
\newcommand{\indOfReg}{r}

\newcommand{\trueP}{\mathcal{P}_{\mathrm{true}}}
\newcommand{\truePsqr}{\mathcal{P}_1}
\newcommand{\truePtrg}{\mathcal{P}_2}

\newcommand{\blind}{0}

\usepackage{xr}
\begin{document}

\def\spacingset#1{\renewcommand{\baselinestretch}%
{#1}\small\normalsize} \spacingset{1}


\if0\blind
{
  \title{\bf Flexible Modeling of Nonstationary Extremal Dependence using Spatially-Fused LASSO and Ridge Penalties}
  \author{Xuanjie Shao$^1$, Arnab Hazra$^2$, Jordan Richards$^3$, and Rapha\"el Huser$^1$\hspace{.2cm}\\
    $^1$Statistics Program, Computer, Electrical and Mathematical Sciences and\\
     Engineering (CEMSE) Division, King Abdullah University of Science \\ and Technology (KAUST), Thuwal 23955-6900, Saudi Arabia\\
    $^2$Department of Mathematics and Statistics, \\Indian Institute of Technology Kanpur, Kanpur 208016, India\\
    $^3$School of Mathematics, University of Edinburgh, Edinburgh EH9 3FD, UK
    }
    \date{}
  \maketitle
} \fi

\if1\blind
{
  \bigskip
  \bigskip
  \bigskip
  \begin{center}
    {\LARGE\bf Flexible Modeling of Nonstationary Extremal Dependence Using Spatially-Fused LASSO and Ridge Penalties}
\end{center}
  \medskip
} \fi


\begin{abstract}
Statistical modeling of a nonstationary spatial extremal dependence structure is challenging. Max-stable processes are common choices for modeling spatially-indexed block maxima, where an assumption of stationarity is usual to make inference feasible. However, this assumption is often unrealistic for data observed over a large or complex domain. We propose a computationally-efficient method for estimating extremal dependence using a globally nonstationary, but locally-stationary, max-stable process by exploiting nonstationary kernel convolutions. We divide the spatial domain into a fine grid of subregions, assign each of them its own dependence parameters, and use LASSO ($L_1$) or ridge ($L_2$) penalties to obtain spatially-smooth parameter estimates. We then develop a novel data-driven algorithm to merge homogeneous neighboring subregions. The algorithm facilitates model parsimony and interpretability. To make our model suitable for high-dimensional data, we exploit a pairwise likelihood to draw inferences and discuss computational and statistical efficiency. An extensive simulation study demonstrates the superior performance of our proposed model and the subregion-merging algorithm over the approaches that either do not model nonstationarity or do not update the domain partition. We apply our proposed method to model monthly maximum temperatures at over 1400 sites in Nepal and the surrounding Himalayan and sub-Himalayan regions; we again observe significant improvements in model fit compared to a stationary process and a nonstationary process without subregion-merging. Furthermore, we demonstrate that the estimated merged partition is interpretable from a geographic perspective and leads to better model diagnostics by adequately reducing the number of subregion-specific parameters.
\end{abstract}

\noindent%

\spacingset{1.5} 
\section{Introduction}

Over the past decades, various parametric max-stable processes (MSPs) have been proposed in the literature \citep[see, e.g.,][]{brown1977extreme, smith1990max, schlather2002models, reich2012hierarchical, opitz2013extremal,beranger2017models} for analyzing spatial extremes, amongst which the Brown--Resnick class is one of the most popular choices. Most classical applications of MSPs assume stationarity and isotropy for simplicity and computational feasibility. Such a simplification is generally unrealistic when we obtain the data over a large or complex geographical domain. Misspecification of the spatial extremal dependence structure can lead to erroneous inferences about spatial risk measures (e.g., quantiles of some spatially aggregated quantities); see \cite{huser2016non} for more details. Therefore, building flexible models capturing nonstationary extremal dependence is crucial.

To capture nonstationary extremal dependence, \cite{blanchet2011spatial} suggest splitting the study domain into distinct subregions and fitting separate stationary extremal dependence models within each subregion, yielding a process that is inconsistent at the subregion boundaries. Locally-stationary processes are also used in the geostatistical literature by \cite{fuentes2001high, fuentes2002interpolation} and \cite{muyskens2022partition}. Using Bayesian information criteria (BIC) or a likelihood ratio test, these authors also propose a subregion-merging procedure to define the boundaries of the local processes. Such models can parsimoniously capture global nonstationarity. Alternatively, the spatial deformation approach, advocated by \cite{sampson1992nonparametric} in the geostatistical literature and by, e.g., \cite{richards2021spatial} in the spatial extremes literature, maps the original spatial sites onto a latent space, where an assumption of stationarity and isotropy is more reasonable. However, the estimation and interpretation of this deformed space can be challenging. A simpler approach is to define the latent space based on covariates, such as climate variables, rather than estimating it \citep{cooley2007bayesian}; this method strongly relies on the choice of covariates for the latent space. Another way of modeling nonstationary extremal dependence is to define a stationary model on an expanded space through multidimensional scaling \citep{chevalier2021modeling}, but computations and interpretation may be difficult.

In the classical spatial setting, \cite{parker2016fused} adopt a local stationarity perspective and propose a computationally-efficient method for modeling global nonstationary spatial dependence using Gaussian processes (GPs). They first partition the domain of interest into a fine grid of subregions, then assign to each its own set of covariance parameters, and finally combine them into a global model through the nonstationary covariance function proposed by \cite{paciorek2006spatial}. Their method fixes the subregion partition and fits an overdetermined model to data, with LASSO regularization utilized to reduce the number of parameters. {\cite{sass2021flexible} also use these penalties to regularize nonstationary marginal parameters in the context of extremes.}

In the spatial extremes setting, \cite{huser2016non} extend the correlation function proposed by \cite{paciorek2006spatial} to incorporate covariates into a nonstationary extremal-$t$ MSP, but their approach is fairly rigid and relies on strong assumptions about the effect of specific covariates on the extremal dependence structure. By contrast, our proposed approach merges ideas from classical geostatistics and extreme value theory by developing a computationally-efficient, and flexible model that can capture global nonstationarity in extremal dependence with locally-stationary subdomains. Specifically, our proposed methodology relies on MSPs and domain partitioning, akin to \cite{parker2016fused}, and the efficient merging of subregions with similar parameter values in a data-driven way. Whilst here we constrain our focus to the Brown--Resnick process, the framework generalizes to other parametric MSPs through replacement of the likelihood function. A regionalization approach to modeling extremal dependence has been adopted previously by \cite{saunders2021regionalisation}; however, they do not consider dependence between subregions. \cite{hector2023distributed} also exploit a domain partitioning strategy for distributed inference with max-stable processes, but {the dependence of their} resulting process model is stationary across the entire spatial domain, and we instead seek to construct a more flexible, nonstationary representation of extremal dependence. Our proposed method also bears similarities with the local likelihood approach advocated by \cite{castro2020local} for modeling threshold exceedances, although we here focus on modeling maxima through max-stable processes instead, and unlike \cite{castro2020local}, our approach takes advantage of the whole dataset to estimate the nonstationary spatial dependence structure. 

The full-likelihood for MSPs is intractable when the number of monitoring sites exceeds $D = 13$ \citep{castruccio2016high, huser2019full}, and the analytical expression of the multivariate density is only available for a small collection of parametric models. Pairwise and triplewise likelihoods are commonly used to draw inference after being studied by \cite{padoan2010likelihood} and \cite{huser2013composite}, with the latter concluding that triplewise likelihood inference has only moderate improvements over the pairwise variant. Alternatively, \cite{huser2022vecchia} adopt the Vecchia likelihood approximation technique \citep{vecchia1988estimation, stein2004approximating}, and several recent works \citep{gerber2021fast, lenzi2021neural, sainsbury2022fast} use artificial neural networks to infer GPs or MSPs. However, both the Vecchia method and neural estimation are impractical when analyzing large spatial datasets. The Vecchia method for spatial extremes is computationally intensive in high dimensions when using composite likelihoods beyond the pairwise case, and a faster estimation technique is necessary for our iterative merging procedure. Neural estimation is also infeasible due to the large number of parameters in our model; we would need to train many neural networks, with different architectures, sequentially due to the nature of the merging algorithm, which would be inefficient. Hence, we prefer to use a pairwise likelihood for inference, which is both simple and quite fast. \cite{huser2016non} suggest a careful choice of a small fraction of observation pairs for achieving computational and statistical efficiency, and we investigate this thoroughly.



The paper is organized as follows. In Section \ref{MaxStableProcesses}, we provide a summary of MSPs and the parametric Brown--Resnick model, along with the details of the pairwise likelihood inference. We detail our methodology and novel contributions, including a nonstationary Brown--Resnick process construction and its inference, in Section~\ref{Methodology}. In Section~\ref{Simulation}, we illustrate the efficacy of our approach with a simulation study, and apply our methodology to gridded monthly maximum temperature data covering the Himalayan and sub-Himalayan regions of Nepal and its surrounding areas in Section~\ref{Application}. Section~\ref{ConcludingRemarks} concludes. 


\section{Max-Stable Processes} \label{MaxStableProcesses}
\subsection{Construction} \label{Construction}
Suppose that $\{X_i(\cdot) : i=1,\ldots,m\}$ are independent and identically distributed (i.i.d.) stochastic processes with continuous sample paths on $\mathcal{S}\subset\mathbb{R}^2$, and there exist sequences of functions $\alpha_{m}(\cdot)>0$ and $\beta_{m}(\cdot)$ defined over $\mathcal{S}$ such that the renormalized process of pointwise maxima $\{\alpha_{m}(\bm{s})^{-1}[\max\{X_1(\bm{s}), \ldots, X_m(\bm{s})\} - \beta_{m}(\bm{s})], \bm{s}\in\mathcal{S} \}$ converges weakly, as $m\rightarrow\infty$, to a process $\{Z(\bm{s}), \bm{s}\in\mathcal{S}\}$ with nondegenerate margins. Then $Z(\cdot)$ is a max-stable process, and for each $\bm{s} \in \mathcal{S}$, $Z(\bm{s})$ follows the generalized extreme value distribution, denoted by $\operatorname{GEV}\{\mu(\bm{s}), \varsigma(\bm{s}), \xi(\bm{s})\}$, with distribution function
\begin{equation}
G_{\bm{s}}(z) = \exp\left( -\left[ 1+\xi(\bm{s})\left\{\dfrac{z-\mu(\bm{s})}{\varsigma(\bm{s})}\right\} \right]^{-1/\xi(\bm{s})}_+ \right),
\label{eq:GEV}
\end{equation}
defined on $\{z \in \mathbb{R}:1+\xi(\bm{s})\{z-\mu(\bm{s})\}/\varsigma(\bm{s})\geq 0\}$, $a_+ = \max(0, a)$, and where $\mu(\bm{s})\in\mathbb{R}$, $\varsigma(\bm{s})>0$, and $\xi(\bm{s})\in\mathbb{R}$ are site-dependent location, scale, and shape parameters, respectively. To separate the dependence structure from the margins, we consider instead the standardized processes $Y_i(\bm{s}) = [1-F_{\bm{s}}\{X_i(\bm{s})\}]^{-1}$, $i = 1,\ldots,m$, where $F_{\bm{s}}(\cdot)$ is the distribution function of $X_i(\bm{s})$. The limiting process $m^{-1}\max\{Y_1(\bm{s}), \ldots, Y_m(\bm{s})\}$, is a (simple) max-stable process with unit Fr\'echet margins, i.e., $\operatorname{GEV}(1,1,1)$. Following \cite{de1984spectral} and \cite{de2007extreme}, under mild regularity conditions, any max-stable process $Z(\cdot)$ with unit Fr\'echet margins can be expressed as
\begin{equation}
    Z(\bm{s}) = \sup_{j\geq 1} W_j(\bm{s})/P_j
    \label{maxstab},
\end{equation}
where for all $j \in \mathbb{N}$, $P_j$ are points of a Poisson process on $(0, \infty)$ with unit rate intensity and $\{W_j(\bm{s}): \bm{s}\in \mathcal{S}\}$ are i.i.d. copies of a non-negative stochastic process satisfying $\mathbb{E}\{W(\bm{s})\} = 1$ for all $\bm{s}\in\mathcal{S}$. The $D$-dimensional joint distribution of $Z(\cdot)$ at sites $\bm{s}_1, \ldots, \bm{s}_D \in \mathcal{S}$ is
\begin{equation}
\operatorname{Pr}\left\{Z\left(\bm{s}_{1}\right) \leq z_{1}, \ldots, Z\left(\bm{s}_{D}\right) \leq z_{D}\right\}=\exp \left\{-V\left(z_{1}, \ldots, z_{D}\right)\right\},
\label{maxsta_cdf}
\end{equation}
where the exponent function $V(\cdot)$ is
\begin{equation}
V\left(z_{1}, \ldots, z_{D}\right)=\mathbb{E}\left[\max \left\{{W\left(\bm{s}_{1}\right)}/{z_{1}}, \ldots, {W\left(\bm{s}_{D}\right)}/{z_{D}}\right\}\right].
\label{exponent}
\end{equation}
The function $V(\cdot)$ has an explicit form only for specific choices of $W(\cdot)$ with one such example arising from the construction of a Brown--Resnick model (see Section~\ref{BrownResnickModel}). 


To measure extremal dependence between $Z(\bm{s}_i)$ and $Z(\bm{s}_j)$, where $i,j \in \{1,\ldots,D\}$, we can use the extremal coefficient $\theta(\bm{s}_i, \bm{s}_j) = V_{(\bm{s}_i, \bm{s}_j)}(1,1)\in [1,2]$ \citep{schlather2003dependence}, where $V_{(\bm{s}_i, \bm{s}_j)}(\cdot, \cdot)$ denotes the restriction of the $D$-variate function $V(\cdot)$ to the variables at sites $\bm{s}_i$ and $\bm{s}_j$ only. The case $1 \leq \theta(\bm{s}_i, \bm{s}_j) < 2$ corresponds to asymptotic dependence with decreasing strength of dependence as $\theta$ increases, and $\theta(\bm{s}_i, \bm{s}_j) = 2$ corresponds to perfect independence between limiting block maxima $Z(\bm{s}_i)$ and $Z(\bm{s}_j)$. The $F$-madogram \citep{cooley2006variograms}, defined by $\nu(\bm{s}_i, \bm{s}_j) = \dfrac{1}{2}\mathbb{E}[|F\{Z(\bm{s}_i)\} - F\{Z(\bm{s}_j)\}|]$ where $F(\cdot)$ here denotes the standard Fréchet distribution function, can be used to estimate the extremal coefficient, as $\theta(\bm{s}_i, \bm{s}_j) = \{ 1+2\nu(\bm{s}_i, \bm{s}_j)\} / \{1-2\nu(\bm{s}_i, \bm{s}_j) \}$. The empirical counterpart $\widetilde{\nu}(\bm{s}_i, \bm{s}_j)$ can be obtained by replacing expectations with sample averages, $F(\cdot)$ with the empirical CDF, and hence the empirical extremal coefficient is defined accordingly.

\subsection{Brown--Resnick Process} \label{BrownResnickModel}

\cite{brown1977extreme} and \cite{kabluchko2009} propose the Brown--Resnick process by specifying $W(\bm{s}) = \exp\{\varepsilon(\bm{s}) - \sigma^2(\bm{s})/2\}$ in (\ref{maxstab}), for a zero-mean Gaussian process $\{\varepsilon(\bm{s}): \bm{s}\in\mathcal{S}\}$ with variance $\sigma^2(\bm{s})$ at site $\bm{s} \in \mathcal{S}$ and semivariogram $\gamma(\cdot, \cdot): \mathcal{S} \times \mathcal{S} \rightarrow [0,\infty)$, i.e., $2\gamma(\bm{s}_i, \bm{s}_j) = \mathbb{E}[\varepsilon(\bm{s}_i) - \varepsilon(\bm{s}_j)]^2$ for any pair of locations $\bm{s}_i, \bm{s}_j \in \mathcal{S}$. For $W(\cdot)$, the bivariate exponent function defined in (\ref{exponent}), for $a = \sqrt{2\gamma(\bm{s}_i, \bm{s}_j)}$, is
\begin{equation} \label{eq:V}
V_{(\bm{s}_i, \bm{s}_j)}\left(z_{i}, z_{j}\right)=\dfrac{1}{z_{i}} \Phi\left\{\dfrac{a}{2}-\dfrac{1}{a} \log \left(\dfrac{z_{i}}{z_{j}}\right)\right\}+\dfrac{1}{z_{j}} \Phi\left\{\dfrac{a}{2}-\dfrac{1}{a} \log \left(\dfrac{z_{j}}{z_{i}}\right)\right\},
\end{equation}
where $\Phi(\cdot)$ is the standard normal distribution function. Here we use the notation $\gamma(\bm{s}_i, \bm{s}_j)$ as the semivariogram may be nonstationary, i.e., a function of both $\bm{s}_i$ and $\bm{s}_j$ rather than only their distance. The theoretical extremal coefficient for the Brown--Resnick model is
\begin{equation}
\theta\left(\bm{s}_i, \bm{s}_j\right)= V_{(\bm{s}_i, \bm{s}_j)}(1,1) =2 \Phi\left[\left\{\gamma(\bm{s}_i, \bm{s}_j) / 2 \right\}^{1 / 2}\right],
\label{eq:extCoef}
\end{equation}
where $\gamma(\bm{s}_i, \bm{s}_j)$ controls the strength and decay of dependence within the process. 


\subsection{Composite Likelihood Inference} \label{CompositeLikelihoodInference}
Full likelihood inference for max-stable processes observed at a large number of sites is difficult, as derivation of the $D$-dimensional joint density involves the differentiation of (\ref{maxsta_cdf}) with respect to $z_1, \ldots, z_D$. {This leads to a summation indexed by all possible partitions of the set $\{1,\ldots,D\}$ with size $D$, i.e., the joint density is a sum of $D$-th Bell number many terms.} The full $D$-dimensional density is computationally intractable even for moderate $D$, e.g., $D>13$ \citep{castruccio2016high}. To tackle this issue, (composite) pairwise likelihood (PL) inference is often used in practice. Denoting $z_{t,i}$ as the $t$-th block maxima recorded at $\bm{s}_i$ for $t=1,\ldots,T$ and $i=1,\ldots,D$, the pairwise log-likelihood $\ell_{PL}$ with $p$-dimensional parameter set $\bm{\psi}$ is
\begin{equation}
\ell_{PL}(\bm{\psi}) = \sum^T_{t=1}\sum_{(i,j)\in\mathcal{O}}\left[ \log\left\{ V_i(z_{t,i}, z_{t,j})V_j(z_{t,i}, z_{t,j}) - V_{ij}(z_{t,i}, z_{t,j})\right\} - V_{(\bm{s}_i, \bm{s}_j)}(z_{t,i}, z_{t,j}) \right],
\label{pairwiseLike}
\end{equation}
where $V_i=\frac{\partial V_{(\bm{s}_i, \bm{s}_j)}}{\partial z_i}$ and $V_{ij}=\frac{\partial^2 V_{(\bm{s}_i, \bm{s}_j)}}{\partial z_i\partial z_j}$, and $\mathcal{O} \subset \mathcal{O}_{total}$ with $\mathcal{O}_{total} = \{(i, j): 1\leq i < j\leq D\}$ taken to be all unique pairs of site indices. If $\mathcal{O} = \mathcal{O}_{total}$, all available observation pairs are utilized and this leads to inefficient inference due to the use of redundant information. Therefore, a careful choice of a significantly smaller number of pairs is suggested by \cite{huser2016non} to achieve computational and statistical efficiency. A possibility is to include a small fraction of strongly dependent pairs \citep{bevilacqua2012estimating}, i.e., the closest ones in the stationary and isotropic case, or using more distant pairs of sites \citep[Chapter 3 of][]{huser2013statistical, huser2014space}. For a nonstationary model, selecting pairs using the lowest pre-computed empirical extremal coefficients, i.e., the ones estimated with the strongest dependence, leads to biases in parameter estimates \citep{huser2016non}. {For nonstationary models, we make use of some alternative methods to sample observation pairs; more details are given in Section~\ref{EfficientInference}.}

Assuming independence between temporal replicates, the maximum pairwise likelihood estimator, denoted $\hat{\bm{\psi}}$, is generally asymptotically Gaussian with convergence rate $\sqrt{T}$. Specifically, if $\bm{\psi}_0$ is the true parameter vector, then, under mild regularity conditions, the distribution of $\hat{\bm{\psi}}$ for large $T$ is $\textrm{Normal}\left(\bm{\psi}_0, \bm{J}(\bm{\psi}_0)^{-1}\bm{K}(\bm{\psi}_0)\bm{J}(\bm{\psi}_0)^{-1} \right)$, where $\bm{J}(\bm{\psi}) = \mathbb{E}\{-\partial^2\ell_{PL}(\bm{\psi})/(\partial\bm{\psi}\partial\bm{\psi}^T)\} \in\mathbb{R}^{p\times p}$ and $\bm{K}(\bm{\psi}) = \operatorname{Var}\{\partial\ell_{PL}(\bm{\psi})/\partial\bm{\psi}\} \in\mathbb{R}^{p\times p}$ are the (pairwise) expected information matrix and variance of the score function, respectively. Evaluating $\bm{J}(\bm{\psi})$ and $\bm{K}(\bm{\psi})$ at an estimate $\hat{\bm{\psi}}$, we can obtain the approximate asymptotic variance of $\hat{\bm{\psi}}$. Model selection in this composite likelihood framework may be performed using either the composite likelihood information criterion (CLIC) or the composite Bayesian information criterion (CBIC), defined by $\operatorname{CLIC} = -2\ell_{PL}(\hat{\bm{\psi}}) + 2\operatorname{tr}\{\bm{J}(\hat{\bm{\psi}})^{-1}\bm{K}(\hat{\bm{\psi}})\}$ and $\operatorname{CBIC} = -2\ell_{PL}(\hat{\bm{\psi}}) + \log(T)\operatorname{tr}\{\bm{J}(\hat{\bm{\psi}})^{-1}\bm{K}(\hat{\bm{\psi}})\}$, respectively, where a model with a lower CLIC or CBIC value is preferred \citep{ng2014model}. 


\section{Methodology} \label{Methodology}

\subsection{Nonstationary Extremal Dependence Structure} \label{NonstationaryExtremalDependenceStructure}

We are interested in modeling nonstationarity in the extremal dependence structure of the Brown--Resnick model, which is determined by its semivariogram $\gamma(\cdot, \cdot)$. As this can be written as a function of the covariance, we build a nonstationary Brown--Resnick process by considering a nonstationary covariance function, rather than directly considering the variogram. The nonstationary convolution-based covariance function of \cite{paciorek2006spatial} is defined by
\begin{equation}
\operatorname{Cov}\{\varepsilon(\bm{s}_i), \varepsilon(\bm{s}_j)\} = \sigma\left(\bm{s}_{i}\right) \sigma\left(\bm{s}_{j}\right) \rho\left(\bm{s}_{i}, \bm{s}_{j}\right),
\label{nonsta-cov}
\end{equation}
with site-dependent sill parameter $\sigma(\bm{s}), \bm{s} \in\mathcal{S}$. Here, $\varepsilon(\cdot)$ denotes a zero-mean Gaussian process, as defined in Section~\ref{BrownResnickModel}. The correlation function $\rho\left(\cdot, \cdot\right): \mathcal{S}\times\mathcal{S}\rightarrow [0,1]$ is 
\begin{equation}
\rho\left(\bm{s}_{i}, \bm{s}_{j}\right)=\left|\bm{\Omega}\left(\bm{s}_{i}\right)\right|^{\frac{1}{4}}\left|\bm{\Omega}\left(\bm{s}_{j}\right)\right|^{\frac{1}{4}}\left|\dfrac{\bm{\Omega}\left(\bm{s}_{i}\right)+\bm{\Omega}\left(\bm{s}_{j}\right)}{2}\right|^{-\frac{1}{2}} \rho^*\left(d_{ij}\right),
\label{nonsta-rho}
\end{equation}
where $\bm{\Omega}\left(\bm{s}\right)$ is a site-dependent $2\times 2$ positive definite kernel matrix controlling the range and shape of local dependence at $\bm{s},$ and $d_{ij}=[(\bm{s}_{i}-\bm{s}_{j})^{T}\{(\bm{\Omega}(\bm{s}_{i})+\bm{\Omega}(\bm{s}_{j}))/2\}^{-1}(\bm{s}_{i}-\bm{s}_{j})]^{1/2}$ is the Mahalanobis distance between sites $\bm{s}_i$ and $\bm{s}_j$. Here $\rho^*(\cdot,\cdot)$ is a stationary isotropic correlation function; we take it to be the commonly-used exponential correlation, $\rho^*(h) = \exp(-h)$ for $h \geq 0$. {Whilst the scaling matrix, $\bm{\Omega}$, can be chosen to permit anisotropic structure in $\rho$, we here follow \cite{parker2016fused} and focus on the locally isotropic case, for which $\bm{\Omega}\left(\bm{s}\right) = \phi(\bm{s})\bm{I}_2$, where $\phi(\bm{s}) \geq 0$ is a site-dependent range parameter, such that $d_{ij}$ is the Euclidean distance between $\bm{s}_i$ and $\bm{s}_j$ rescaled by $[\{\phi(\bm{s}_i)+\phi(\bm{s}_j)\}/2]^{1/2}$}. More complex non-isotropic cases are possible with general forms of $\bm{\Omega}(\bm{s})$;  one such example would be obtained by allowing the diagonal elements of $\bm{\Omega}(\bm{s})$ to be unequal or the off-diagonal elements to be nonzero. Substituting (\ref{nonsta-cov}) {into $2\gamma(\bm{s}_i, \bm{s}_j) = \mathbb{E}[\varepsilon(\bm{s}_i) - \varepsilon(\bm{s}_j)]^2$} yields the nonstationary exponential variogram 
\begin{equation}\label{eq:ExpVario}
2\gamma(\bm{s}_i, \bm{s}_j) = \sigma^2\left(\bm{s}_{i}\right) + \sigma^2\left(\bm{s}_{j}\right) - 2\sigma\left(\bm{s}_{i}\right) \sigma\left(\bm{s}_{j}\right) \rho\left(\bm{s}_{i}, \bm{s}_{j}\right), 
\end{equation}
which can be used in \eqref{eq:V} for constructing nonstationary Brown--Resnick processes.

\subsection{Sampling Schemes for Efficient Inference} \label{EfficientInference}
To reduce redundant information and balance the computational and statistical efficiency required in optimizing the composite likelihood mentioned in Section~\ref{CompositeLikelihoodInference}, we investigate two strategies for inference: simple and stratified sampling. These strategies determine how to select the set of paired sites, $\mathcal{O}$. Our simple strategy is random sampling of a small fraction of observation pairs with uniform probability. With stratified sampling, we collect observation pairs in predefined distance classes, and then draw the same percentage of pairs from each class; this approach ensures that the distribution of distances among sampled pairs is approximately uniform, and so distant pairs have a similar \enquote{weight} in the pairwise likelihood as close pairs. Our simulation study in Section \ref{SelectObservationPairs} investigates the statistical efficiency of both sampling schemes.

\subsection{Domain Partitioning and Dependence Regularization} \label{DomainPartition}
For a fully over-determined model, each site would have its own variogram parameters, and hence the number of parameters would be proportional to the number of monitoring sites, $D$. Such a large number of unique parameters may not be required to represent the extremal dependence structure of a process reasonably, and in fact, their uniqueness at each site may be unrealistic in applications where we can reasonably assume homogeneity of parameters at nearby sites. To improve model realism and parsimony, we use a data-driven approach to regularize the dependence and enforce equality of parameters for sites within a subregion of the domain. 

To do so, we first partition the whole domain into a partition $\mathcal{P}$ which consists of $\numOfReg \leq D$ subregions $\{\mathcal{R}_{\indOfReg}\}^{\numOfReg}_{\indOfReg=1}$ and assign each subregion its own dependence parameters. By an abuse of notation, the vector of sill parameters reduces to $\bm{\sigma}^2 := \{\sigma^2_1, \ldots, \sigma^2_{\numOfReg}\}^T$, and $\sigma^2(\bm{s}_i) = \sigma^2_{\indOfReg}$ if $\bm{s}_i \in \mathcal{R}_{\indOfReg}$, for $\indOfReg = 1,\ldots,\numOfReg$, $i = 1,\ldots,D$, and similarly for the range parameter vector $\bm{\phi}$. This partition should be constructed to balance parameter estimation accuracy and model flexibility. If subregions are too small, estimation of the corresponding parameters can be numerically unstable; if subregions are too large, the model may be incapable of capturing fine-scale nonstationarity in the extremal dependence structure. Similar to \cite{parker2016fused}, we also impose a penalty to regularize the dependence parameters and fit the model by maximizing the penalized pairwise log-likelihood (PPL), given by 
\begin{equation}
\ell_{PPL}(\bm{\psi}) = \ell_{PL}(\bm{\psi}) - \sum_{i=1}^{2} \lambda_{i} \sum_{\indOfReg_{1} \sim \indOfReg_{2}}\left|\psi_{i \indOfReg_{1}}-\psi_{i \indOfReg_{2}}\right|^{q},
\label{penPair}
\end{equation}
where the tuning parameter $\eleOfDG_i$ control the similarity between parameters for neighboring subregions, $\psi_{1\indOfReg} = \log(\sigma^2_{\indOfReg})$ and $\psi_{2\indOfReg} = \log(\phi_{\indOfReg})$ are the log-sill and log-range parameters, respectively, for subregion $\mathcal{R}_{\indOfReg}$, and the operation $\indOfReg_1\sim \indOfReg_2$ refers to neighboring subregions $\mathcal{R}_{\indOfReg_1}$ and $\mathcal{R}_{\indOfReg_2}$ sharing a boundary. Here \eqref{penPair} with $q = 1$ and $q = 2$ correspond to LASSO ($L_1$) and ridge ($L_2$) regularization, respectively. While the former penalizes small discrepancies (and can enforce parameters for neighboring subregions to be exactly equal), the latter penalizes larger discrepancies and tends to produce smoother parameter surfaces. 



\subsection{Partition Update} \label{PartitionUpdate}
Instead of fixing the domain partition, as in \cite{parker2016fused}, we iteratively update the partition $\mathcal{P}$ and reduce the number of parameters required for our model. Alongside increased parsimony, this approach allows us to better identify stationary subdomains of $\mathcal{S}$, which also leads to easier model interpretation. We achieve this in a data-driven way, without the need for domain knowledge or covariate information. Our strategy is to construct a base partition $\mathcal{P}^{(0)}$ via a random partitioning of the domain into fine subregions of similar shapes and sizes. We can achieve this by either creating a regular grid of subregions, or by applying a clustering algorithm directly to the sites and partitioning the study domain into similarly-shaped convex polygons containing observational sites. We then iteratively update the partition by merging neighboring subregions, according to some quantifiable improvement in the model fit. {We describe now a general overview of the algorithm; full details of the algorithm are to follow.}

{With a base partition $\mathcal{P}^{(0)} = \{\mathcal{R}_{\indOfReg}\}_{\indOfReg=1}^{\numOfReg^{(0)}}$ comprising $\numOfReg^{(0)}$ subregions with $1 <~\numOfReg^{(0)} \leq~D$, our target is to estimate a merged partition $\hat{\mathcal{P}}$. For iterations $j=1,2,\ldots$, a new tentative partition $\mathcal{P}^{(j)}$ of size $\numOfReg^{(j)} < \numOfReg^{(j-1)}$ is created by merging ``similar'' neighboring subregions in partition $\mathcal{P}^{(j-1)}$. We characterize similarity between subregions through differences in the dependence parameter estimates; if these differences are smaller than a threshold $\eleOfThr$, belonging to a set of candidate thresholds $\colOfThr$, we merge the corresponding subregions to create a tentative partition and accept this partition as the current best estimate $\mathcal{P}^{(j)}$ if it improves the model fit. To smooth parameter estimates, we impose fused-LASSO or -ridge penalization, with penalty $\bm{\lambda}$ (this is tuned via our $\bm{\lambda}$-tuning procedure, which is described by the function \texttt{LambdaTuning} in Algorithm~\ref{alg:comb}). The penalty $\bm{\lambda}$ belongs to a set of penalty parameters $\bm{\Lambda}$, which we iteratively augment throughout the partition update algorithm (see \texttt{UpdateGrid} in Algorithm~\ref{alg:comb}). Specific details are to follow, and the pseudo-code of the full procedure for estimating a partition $\hat{\mathcal{P}}$ is detailed in Algorithm~\ref{alg:comb}.} 

\begin{algorithm}[t!]
\caption{Partition update}\label{alg:comb}
\begin{algorithmic}[1]
\Require Base partition $\mathcal{P}^{(0)} = \{\mathcal{R}_{1}, \ldots, \mathcal{R}_{\numOfReg^{(0)}}\}$, descending grid of penalty parameters $\colOfDG$, a holdout set $\bm{z}_{\mathrm{hold}}$, initial penalty $\bm{\lambda}=(\infty, \infty)^T$.
\Ensure A final partition $\hat{\mathcal{P}}$ of size $\hat{\numOfReg}$, dependence parameter estimates $\hat{\bm{\psi}}$, and smoothing penalty $\hat{\bm{\lambda}}$.
\State $\hat{\mathcal{P}} \gets \mathcal{P}^{(0)}, \hat{\numOfReg} = \numOfReg^{(0)}$
\State $\hat{\bm{\psi}}, \hat{\bm{\lambda}} \gets$ \texttt{LambdaTuning}($\mathcal{P}^{(0)}$, $\colOfDG$, $\bm{\lambda}$) \Comment{tune $\bm{\lambda}$ on the base partition}
\State Calculate distances $\widetilde{\colOfThr} = \{d(\hat{\bm{\psi}}_{r_{1}}, \hat{\bm{\psi}}_{r_{2}}) : r_{1} \sim r_{2}, r_{1},r_{2}=1,\ldots,\hat{\numOfReg}\}$
\State Take $q_{\tau}$ as empirical $\tau$-quantile of $\widetilde{\colOfThr}$
\State Choose $\colOfThr = \{\eleOfThr \in \widetilde{\colOfThr}: \eleOfThr < q_{\tau}\}$, \Comment{candidates for $\eleOfThr$}
\State Sort $\colOfThr := \{\eta_h\}_{h=1}^H$ decreasingly, with $H := |\colOfThr|$ 
\State Set $\indOfThr \gets 1$
\While{$\indOfThr \leq \numOfThr$} \Comment{continue if not all $\eleOfThr\in\colOfThr$ are tested}
    \State $\colOfDG \gets$ \texttt{UpdateGrid}($\hat{\bm{\lambda}}$, $\bm{\lambda}$, $\colOfDG$) \Comment{augment $\colOfDG$}
    \Repeat \Comment{test each $\eleOfThr$ in $\colOfThr$}
        \State $\rvOfThr \gets \eleOfThr_{\indOfThr}$ \Comment{$\indOfThr$-th largest $\eleOfThr$ in $\colOfThr$}
        \State $\mathcal{P}^*, \numOfReg^* \gets$ \texttt{MergeSubregion}($\hat{\mathcal{P}}$, $\hat{\bm{\psi}}$, $\rvOfThr$) \Comment{obtain a tentative partition and its size}
        \State $\hat{\bm{\psi}}^*, \hat{\bm{\lambda}}^* \gets$ \texttt{LambdaTuning}($\mathcal{P}^*$, $\colOfDG$, $\hat{\bm{\lambda}}$) \Comment{tune $\bm{\lambda}$ on the tentative partition}
        \State $\indOfThr \gets \indOfThr + 1$
    \Until{$\ell_{PPL}(\hat{\bm{\psi}}^*; \bm{z}_{\mathrm{hold}}) > \ell_{PPL}(\hat{\bm{\psi}}; \bm{z}_{\mathrm{hold}})$ or $\indOfThr > \numOfThr$} \Comment{if an increase in holdout PPL}
    \If{$\indOfThr\leq \numOfThr$} \Comment{accept the tentative partition}
        \State Update $\hat{\bm{\psi}} \gets \hat{\bm{\psi}}^*$; $\hat{\bm{\lambda}} \gets \hat{\bm{\lambda}}^*$; $\hat{\mathcal{P}} \gets \mathcal{P}^*$; $\hat{\numOfReg} \gets \numOfReg^*$
        \State Update $\widetilde{\colOfThr}$, $q_{\tau}$, and $\colOfThr$ accordingly as on lines 3--6
        \State Reset $\indOfThr \gets 1$
    \EndIf
\EndWhile
\end{algorithmic}
\end{algorithm}

To choose {``similar''} subregions to merge at iteration $j$ of Algorithm~\ref{alg:comb}, \cite{muyskens2022partition} suggest randomly sampling neighboring pairs from all available neighboring subregions, and testing their merge; this is performed exhaustively, without replacement, until some acceptance criterion is satisfied. Different criteria {for assessing model fit} have been proposed in the literature, e.g., BIC \citep{fuentes2001high} or deviance \citep{muyskens2022partition}, but these are not applicable when the likelihood is penalized or composite, as is the case for our model. Instead, we {quantify model fit and} select a partition using the penalized pairwise log-likelihood, denoted by $\ell_{PPL}(\hat{\bm{\psi}}; \bm{z}_{\mathrm{hold}})$, which we compute over a holdout set $\bm{z}_{\mathrm{hold}}$; note that the estimate $\hat{\bm{\psi}}$ maximizes $\ell_{PPL}$ on a training set $\bm{z}_{\mathrm{train}}$. We construct $\bm{z}_{\mathrm{hold}}$ by first defining a set of holdout sites and then holding out all observations at said sites; the remaining data comprises $\bm{z}_{\mathrm{train}}$. The penalty term in (\ref{penPair}) inherently includes information about the number of parameters in the model; the total number of neighboring subregion pairs decreases whenever subregions merge and hence the number of model parameters also decreases. Selecting the model with the larger penalized pairwise log-likelihood balances model flexibility and parsimony.

{An exhaustive testing of merges for every iteration of our Algorithm~\ref{alg:comb}} invokes a high computational burden, and thus we instead choose to merge only neighboring subregions where the $L_1$ distance between their parameter estimates, denoted by $d(\hat{\bm{\psi}}_{r_{1}}, \hat{\bm{\psi}}_{r_{2}})$ for $r_1, r_2=1,\dots,\numOfReg^{(j-1)}$, is smaller than some predefined threshold $\rvOfThr>0$, which may change with iteration $j$. {Specifically, we merge all neighboring subregions $\mathcal{R}_{1}, \mathcal{R}_{2}\in \mathcal{P}^{(j-1)}$ if $d(\hat{\bm{\psi}}_{r_{1}}, \hat{\bm{\psi}}_{r_{2}}) < \rvOfThr$, which generates a tentative partition $\mathcal{P}^* = \{\mathcal{R}_i\}_{i=1}^{\numOfReg^*}$ for iteration $j$, where ${\numOfReg^*=|\mathcal{P}^*|\leq \numOfReg^{(j-1)}}$.} This procedure{, denoted by \texttt{MergeSubregion} in Algorithm~\ref{alg:comb},} allows us to reduce computational time by merging multiple subregions simultaneously. 

Intuitively, parameter estimates for neighboring subregions that should be merged are closer in value than those of neighboring subregions that should not be merged. However, the merged partition may be over-merged and hence lacking in flexibility if $\rvOfThr$ is too large, while the methodology suffers from computational inefficiency if $\rvOfThr$ is too small and few subregions get merged in each iteration. To determine the threshold $\rvOfThr$ for each partition, we propose choosing a set of descending candidate thresholds according to all differences in parameter estimates for neighboring subregions in partition $\mathcal{P}^{(j-1)}$; we denote the differences by $\widetilde{\colOfThr}$. For $\tau \in (0,1]$, we take the set of threshold candidates $\colOfThr$ as elements of $\widetilde{\colOfThr}$ {below} the empirical $\tau$-quantile of $\widetilde{\colOfThr}$, with $H := |\colOfThr|$. Then, for components $\rvOfThr\in\colOfThr$ ordered from largest to smallest, we sequentially test the merge of {all neighboring subregions with paired distance of estimates less than $\eleOfThr$}, until we observe an increase in $\ell_{PPL}(\hat{\bm{\psi}}; \bm{z}_{\mathrm{hold}})$. If we do not observe any increase using the smallest candidate threshold in $\colOfThr$,  Algorithm~\ref{alg:comb} terminates and we retain the last accepted partition, i.e., $\hat{\mathcal{P}} = \mathcal{P}^{(j-1)}$. Here $\tau$ can be viewed as the step length for the optimization search of Algorithm~\ref{alg:comb}. In practice, $\tau$ can be either a pre-specified hyperparameter or adaptive to the current partition. {The latter can be achieved by clustering all estimated differences into two clusters (small and large) and defining $\tau$ as the relative size of the small cluster, i.e., the proportion of the size of the small cluster to that of the whole}. This choice is motivated by the intuition that we can group neighboring subregion pairs into two classes: the pairs that should be merged, i.e., those with smaller differences in parameter estimates, and those that should not.

{We allow the smoothing penalty $\bm{\lambda}$ to vary across iterations of Algorithm~\ref{alg:comb}, via our $\bm{\lambda}$-tuning procedure,} using a similar approach to \cite{parker2016fused}. For the initial partition $\mathcal{P}^{(0)}$, we begin with the fully stationary model, i.e., with $\lambda_1 = \lambda_2 = \infty$, and then iteratively fit two competing models with new {values of $\bm{\lambda}$. For example, in the first iteration,} two different models are fitted, with $(\eleOfDG_{1,2}, \infty)$ and with $(\infty, \eleOfDG_{2,2})$, {where finite $\lambda_{1,2},\lambda_{2,2}>0$ are chosen a priori.} If a larger $\ell_{PPL}(\hat{\bm{\psi}}; \bm{z}_{\mathrm{hold}})$ is achieved by one of them, we choose the model with the largest improvement. {We iteratively decrease the values of $\bm{\lambda}$ according to candidate penalty parameter sets $\colOfDG_i = \{\infty, \eleOfDG_{i,2}, \ldots, \eleOfDG_{i,\numOfDG_i}\}$ for $i = 1,2$. We then fit two new similarly competing models and} repeat this procedure until no improvement is observed (see \texttt{LambdaTuning} in Algorithm~\ref{alg:comb}), {leading to an optimal smoothing penalty $\hat{\bm{\lambda}}$ and corresponding dependence parameter estimates $\hat{\bm{\psi}}$.} 

The smoothing penalty $\bm{\lambda}$ determines model complexity and, thus, should vary with the partition $\mathcal{P}$. For every new tentative partition $\mathcal P^*$, we tune $\bm{\lambda}$ to get an optimal value $\hat{\bm{\lambda}}^*$. Using $\hat{\bm{\lambda}}^*$ and the corresponding dependence parameter estimates $\hat{\bm{\psi}}^*$ over the training set, we calculate $\ell_{PPL}(\hat{\bm{\psi}}^*; \bm{z}_{\mathrm{hold}})$ and use this for partition selection. The optimal $\hat{\bm{\lambda}}^*$ is sensitive to the resolution of the grids $\colOfDG_i, i=1,2$, and thus we allow these to change with iteration $j$ as the optimal partition is updated, relative to the previous value of $\hat{\bm{\lambda}}^*$. {We thus extend on the approach by \cite{parker2016fused} and allow $\bm{\Lambda} = (\bm{\Lambda}_1, \bm{\Lambda}_2)$ to vary across iterations, via an augmentation scheme.} The penalty parameter sets, $\bm{\Lambda}_i,i=1,2,$ are chosen to be descending grids to ensure that we gradually lessen the penalty to fit the model better (in terms of pairwise likelihood) while smoothing the parameter estimates. We initially choose $\bm{\Lambda}_i$ as the descending sequence of the powers of a base number (in our case, 2). However, the initial choice of $\bm{\Lambda}_i$ is somewhat arbitrary, as we also adapt the tuning parameter candidates during the $\bm{\lambda}$-tuning procedure; we achieve this by introducing intermediate values to $\bm{\Lambda}_i$ if the $\bm{\lambda}$-tuning procedure does not produce a new optimal value of $\bm{\lambda}$ in the previous iteration of {Algorithm~\ref{alg:comb}}. For example, in our simulation study and application, we take both initial grids, $\colOfDG_1$ and $\colOfDG_2$, to be $\{\infty, 2^5, 2^4, 2^3, 2^2, 2^1, 2^0, 2^{-1}\}$, and the initial smoothing penalty vector as $\bm{\lambda} = \{\infty, \infty\}$. For the $j$-th iteration of {Algorithm~\ref{alg:comb}}, the $\bm{\lambda}$-tuning procedure takes the optimal penalty parameters in step $j-1$, denoted by $\hat{\bm{\lambda}}$, as the initial value in the current iteration, i.e., $\bm{\lambda}^* := \hat{\bm{\lambda}} = (\hat{\lambda}_1, \hat{\lambda}_2)$. Then, the descending candidate grids used for iteration $j$ are taken to be invariant if $\hat{\lambda}_i$ changes with the $\bm{\lambda}$-tuning procedure in iteration $j-1$. Otherwise, as $\lambda_i$ may be over-tuned, we increase the resolution of $\colOfDG_i$ by adding the average of pairs of values adjacent to $\hat{\bm{\lambda}}$. That is, if $\hat{\lambda}_i = \eleOfDG_{i,\indOfDG}$, we add $(\eleOfDG_{i,\indOfDG} + \eleOfDG_{i,\indOfDG+1})/2$ and $(\eleOfDG_{i,\indOfDG} + \eleOfDG_{i,\indOfDG-1})/2$ to $\colOfDG_i$. This procedure ``smooths'' the descending grid and extends the optimization space for $\bm{\lambda}$. The update step for $\bm{\Lambda}$ is \texttt{UpdateGrid} in Algorithm \ref{alg:comb}.

The number of subregions in the final partition $\hat{\mathcal{P}}$ is another problem of interest. As we take a data-driven approach, the final partition $\hat{\mathcal{P}}$ may be sensitive to the choice of holdout set and observation pairs used for optimization. Hence, we advocate estimating multiple different $\hat{\mathcal{P}}$ using different holdout sets and observation pairs and selecting the most reasonable one.


\section{Simulation Study} \label{Simulation}
\subsection{Overview}\label{Overview}
In the simulation study, we first investigate the statistical properties of the maximum pairwise log-likelihood estimator using different proportions of pairs for inference in Section \ref{SelectObservationPairs}. We then study the model performance on simulated data in terms of model fit and true partition recovery in Sections \ref{ModelPerformance} and \ref{TruePartitionRecovery}, respectively.

\begin{figure}[hbt!]
    \centering
    \includegraphics[height = 0.22\linewidth]{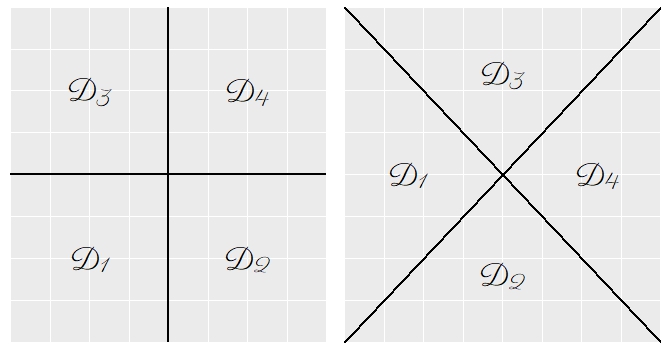}
    \caption{True partitions for the simulation study. Left: true partition $\truePsqr$, square grid; right: true partition $\truePtrg$, triangular grid.}
    \label{pic:4truePar}
\end{figure}

We generate data with unit-Fr\'echet margins on a $\mathcal{S} = 40\times 40$ grid of sites within the unit square $\mathcal{D} = [0,1]^2$, thus with $D=1600$ sites. We then divide $\mathcal{D}$ into four subdomains, $\mathcal{D}_r, r = 1,\ldots,4$, under two partition choices: a square grid (true partition $\truePsqr$) or a triangular grid (true partition $\truePtrg$), see Figure \ref{pic:4truePar}. Each subdomain has its own sill $\sigma^2$ and range parameter $\phi$, as defined in (\ref{nonsta-rho}). We consider two different cases for the extremal dependence structure and enforce that only one of the two parameters varies between subregions. The other remains fixed over $\mathcal{D} = \bigcup^4_{r=1}\mathcal{D}_r$. These two cases (shown below) can help us to understand and compare the influence of each parameter on the extremal dependence structure.
\begin{itemize}
    \item Case C1: $\sigma^2$ varies with values $\{0.5, 2, 2, 8\}$ for subdomains $\mathcal{D}_1$, $\mathcal{D}_2$, $\mathcal{D}_3$, and $\mathcal{D}_4$, respectively, with $\phi = 2$ fixed over $\mathcal{D}$.
    \item Case C2: $\phi$ varies with values $\{0.05, 0.2, 0.2, 0.8\}$ for subdomains $\mathcal{D}_1$, $\mathcal{D}_2$, $\mathcal{D}_3$, and $\mathcal{D}_4$, respectively, with $\sigma^2 = 0.2$ fixed over $\mathcal{D}$.
\end{itemize}

For each case, we then generate approximate samples from our nonstationary Brown--Resnick process. Here, we simulate $T = 100$ independent replicates of $Z(\cdot)$ and repeat the experiment $N = 100$ times to assess the performance of our estimation approach. The initial descending grid for the $\bm{\lambda}$-tuning procedure is $\colOfDG_i = (\infty, 2^5, 2^4, 2^3, 2^2, 2^1, 2^0, 2^{-1},$ $2^{-2},0), i=1,2$.

\subsection{Selecting Observation Pairs} \label{SelectObservationPairs}

Suppose $\hat{\bm{\psi}}_{\mathcal{O}} = \max_{{\bm{\psi}}_{\mathcal{O}}}\ell_{PL}(\bm{\psi}; \mathcal{O})$ is the maximum pairwise likelihood estimator, evaluated using the subset of observation pairs $\mathcal{O} \subset \mathcal{O}_{\mathrm{total}}$, as described in Section \ref{CompositeLikelihoodInference}. Using different percentages of observation pairs, varying from $0.005\%$ to $100\%$, and nonstationary data simulated according to the scheme given in Section \ref{Overview}, we estimate $\hat{\bm{\psi}}_{\mathcal{O}}$ to study the variability of the estimator around the true parameter values. We select the observation pairs with the methods discussed in Section \ref{Methodology}, i.e., simple and stratified random sampling, with the latter using ten equal-length distance classes. We fit nonstationary models with true partitions $\truePsqr$ and $\truePtrg$, and we consider both cases of parameter specification C1 and C2; results for true partition $\truePtrg$ appear to be very similar to those for $\truePsqr$, and hence we omit these results for brevity. We quantify the estimation accuracy through the root mean squared error (RMSE), e.g., for case C1, we compute $\operatorname{RMSE}(\hat{\bm{\sigma}}^2) = \sqrt{\frac{1}{4}\sum^4_{r=1}(\hat{\sigma}^2_{r} - \sigma^2_{r})^2}$ for all $N$ experiments, and similarly for C2 and the range parameter $\phi$.
\begin{figure}[t!]
    \centering
    \subfloat{{\includegraphics[width=0.98\linewidth]{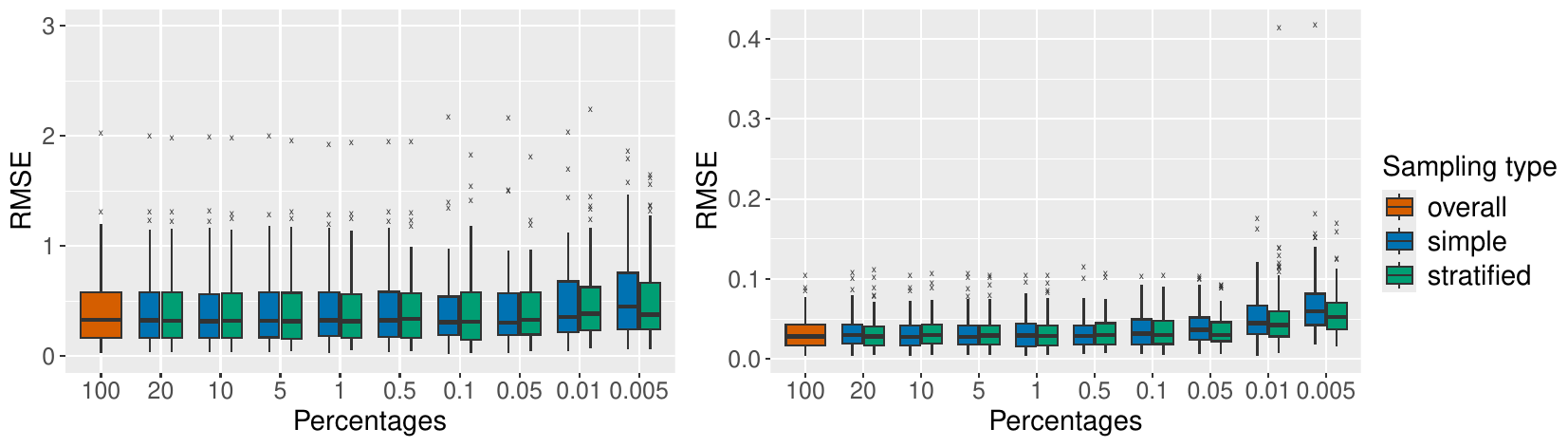} }}
    \caption{Boxplots of the RMSE for $\hat{\bm{\psi}}_{\mathcal{O}}$ using all observation pairs (orange), the simple random sampling (blue), and stratified sampling (green) schemes, under case C1, i.e., varying $\sigma^2$ (left) and, C2, i.e., varying $\phi$ (right).}
    \label{pic:percentageOfPairs-nonsta}
\end{figure}

The left and right panels of Figure \ref{pic:percentageOfPairs-nonsta} report the RMSE boxplots for cases C1 and C2, i.e., $\operatorname{RMSE}(\hat{\bm{\sigma}}^2)$ and $\operatorname{RMSE}(\hat{\bm{\phi}})$, respectively. We do not observe any significant difference in the performance of the sampling strategies if more than $0.05\%$ of observation pairs are used for fitting, while the optimization time decreases proportionally; more details are in Table~\ref{tab:timeandmeanrmse} in the Supplementary Material. However, the stratified sampling scheme tends to slightly outperform the simple one when {few ($0.005\%$)} pairs are used, and usually provides less biased estimates in both cases C1 and C2, especially for the estimation of the range parameter; {see Table~\ref{tab:timeandmeanrmse} in the Supplementary Material for detailed RMSE values. Moreover, the computation times are similar (see Table \ref{tab:timeandmeanrmse}). We therefore advocate using stratified sampling.} In our application, we use $0.1\%$ of observation pairs for fitting, to balance computational and statistical efficiency. We use simple random sampling in Section \ref{ModelPerformance}, but in our real data application in Section \ref{Application}, we use stratified sampling.



\subsection{Model Performance} \label{ModelPerformance}

We now compare the performance of three extremal dependence models: a fully stationary Brown--Resnick (BR) model (model~S), a nonstationary BR model using the base partition (Model~B), and a nonstationary BR model using the merged partition (Model~M). For Model B and Model M, we consider both $L_1$ and $L_2$ regularization to ``smooth'' parameter estimates. We simulate data for both settings of the extremal dependence parameters, i.e., C1 and C2, on both true partitions, $\truePsqr$ and $\truePtrg$. 

To study the influence of base partition $\mathcal{P}^{(0)}$ misspecification (with respect to the true partition) on the final model fit, we consider three scenarios: one scenario where the base partition $\mathcal{P}^{(0)}$ is well-specified and it is feasible to obtain $\hat{\mathcal{P}}=\trueP$, for true partition $\trueP = \truePsqr$, and two scenarios where $\mathcal{P}^{(0)}$ is misspecified, i.e., $\hat{\mathcal{P}}=\trueP$ is infeasible, and $\trueP = \truePsqr$ or $\truePtrg$, respectively. We construct the base partition from $\numOfReg^{(0)}=100$ subregions under all three scenarios. For true partition $\truePsqr$, i.e., the square grid, we construct the well-specified $\mathcal{P}^{(0)}$ by considering a regular subgrid, whereas the misspecified $\mathcal{P}^{(0)}$ is generated as the convex hull of the clusters created by applying $k$-means clustering to the spatial coordinates, and where some subregions overlap the boundaries of $\trueP$. For true partition $\truePtrg$, we can only consider a misspecified base partition; we construct this similarly using $k$-means clustering. Among the $D = 1600$ observation sites, $15\%$ are randomly selected as a validation set ($D_{\mathrm{valid}} = 240$). For the nonstationary models, we randomly select another $15\%$ of the sites as a holdout set ($D_{\mathrm{hold}} = 240$) while ensuring that at least one site is within each subregion in the base partition. 

To evaluate parameter estimation, we adopt the integrated RMSE, e.g., for the sill estimates $\hat{\bm{\sigma}}^2$, we have $\operatorname{IntRMSE}(\hat{\bm{\sigma}}^2) = \sqrt{\frac{1}{N}\sum^N_{n=1}\frac{1}{D}\sum^D_{j=1}\left(\hat{\sigma}^2_n(\bm{s}_j) - \sigma^2(\bm{s}_j)\right)^2},$ where $\sigma^2(\bm{s}_j)$ denotes the true parameter value and $\hat{\sigma}_n^2(\bm{s}_j)$ denotes the fitted value for the $n$-th experiment at site $\bm{s}_j$. To assess model fits, we compute, for each model, the pairwise log-likelihood on the validation set, the penalized pairwise log-likelihood on the holdout set, and the CLIC and CBIC on the training set. {Likelihood-based measures are computed using bootstrap with 1000 samples. For each trial, the training, holdout, and validation sets are randomly generated.} Results are given in Table~\ref{tab:extVarySill}, and Tables~\ref{tab:extVaryRange} and~\ref{tab:triangle} in the Supplementary Material.

\begin{table}[t!]
    \centering
    \caption{Performance of Models S, B, and M for case C1 (varying $\sigma^2$) with true partition $\truePsqr$. The relative factors shown in subscript parentheses highlight the ratios between the nonstationary and stationary models. In the first four rows, the value of the best model has been subtracted. The best value in each row is in bold.}
    \resizebox{\columnwidth}{!}{%
    \begin{tabular}{l|l|ll|ll|ll|ll}
    \hline
    & Model S & \multicolumn{8}{c}{Nonstationary}\\
    \cline{3-10}
    & & \multicolumn{4}{c|}{Model~B} & \multicolumn{4}{c}{Model~M}\\
    \cline{3-10}
    & & \multicolumn{2}{c|}{well-specified} & \multicolumn{2}{c|}{misspecified} & \multicolumn{2}{c|}{well-specified} & \multicolumn{2}{c}{misspecified}\\
    \cline{3-10}
    & & $L_1$ & $L_2$ & $L_1$ & $L_2$ & $L_1$ & $L_2$ & $L_1$ & $L_2$\\
    \hline
    PL Diff. & $-120749$ & $-300$ & $-363$ & $-3927$ & $-7287$ & $\bm{0}$ & $-16$ & $-4156$ & $-7443$ \\
    PPL Diff. & -- & $-1317$ & $-1076$ & $-12222$ & $-14960$ & $-55$ & $\bm{0}$ & $-9832$ & $-12993$ \\
    CLIC Diff. & $7605397$ & $655182$ & $378508$ & $1293988$ & $1555086$ & $10109$ & $\bm{0}$ & $1038650$ & $1240018$ \\
    CBIC Diff. & $7521439$ & $1499224$ & $852815$ & $1678787$ & $1981966$ & $24031$ & $\bm{0}$ & $1046065$ & $1256174$ \\
    IntRMSE $\hat{\bm{\sigma}}^2$ & $2.92_{(1)}$ & $0.82_{(0.28)}$ & $0.83_{(0.28)}$ & $1.17_{(0.4)}$ & $1.17_{(0.4)}$ & $\bm{0.53_{(0.18)}}$ & $0.58_{(0.2)}$ & $1.09_{(0.37)}$ & $1.1_{(0.38)}$ \\
    IntRMSE $\hat{\bm{\phi}}$ & $\bm{0.04_{(1)}}$ & $0.15_{(3.48)}$ & $0.15_{(3.47)}$ & $0.14_{(3.18)}$ & $0.14_{(3.14)}$ & $0.05_{(1.08)}$ & $0.06_{(1.38)}$ & $0.11_{(2.56)}$ & $0.1_{(2.34)}$ \\
    \hline
    \end{tabular}}
    \label{tab:extVarySill}
\end{table}


When only the sill varies (Table \ref{tab:extVarySill}), all nonstationary models exhibit larger estimates of the pairwise and penalized pairwise log-likelihoods, as well as lower estimates of the CLIC and CBIC, when compared to the stationary model. This ordering indicates that the nonstationary models provide better fits (as expected). Model~M outperforms Model~B as we observe larger penalized pairwise log-likelihood and lower CLIC and CBIC values. The nonstationary models provide a more accurate estimation of the spatially-varying $\sigma^2$ (with both penalties), observable through the reduced integrated RMSE. However, for $\phi$ estimation, some estimation accuracy is sacrificed when using nonstationary models (approximately 3--3.5 times integrated RMSE for Model~B, and only about 1--2.5 times for Model~M, compared to the stationary model); however, from a practical standpoint, these errors are not large. This may be because poor estimation of $\phi$ does not strongly influence estimates of the overall spatial extremal dependence. A similar phenomenon is present in classical geostatistics when estimating the range parameter of a Gaussian process \citep{stein1999interpolation}. 

It is noteworthy that nonstationary models with a misspecified base partition outperform the stationary model, which validates the efficacy of nonstationary models. We further note that Model~M provides lower integrated RMSE than Model~B for both the sill and range parameters, which indicates that merging subregions improves parameter estimation. {For likelihood-based measures, $L_2$ regularization performs slightly better for models with a well-specified base partition, while $L_1$ regularization generally outperforms $L_2$ in a more realistic scenario, i.e., where the base partition is misspecified, especially for Model~M. Both regularizations are comparable in terms of parameter estimation accuracy.} As expected, nonstationary models with a well-specified partition have better performance. {Tables \ref{tab:extVaryRange} and \ref{tab:triangle} in the Supplementary Material show the results for case C2 of the extremal dependence structure (varying $\phi$) with a true partition $\trueP = \truePsqr$ and both dependence cases (C1 and C2) with a true partition $\trueP = \truePtrg$, respectively.} Similarly to case C1 with true partition $\trueP = \truePsqr$, we again observe that nonstationary models show their advantages in terms of both model fit and estimation of spatially-varying parameters, and Model~M further generally outperforms Model~B. We sacrifice some accuracy for the fixed parameter estimation when using nonstationary models, but Model~M, compared to Model~B, demonstrates the ability to reduce the estimation error. {In terms of Model~M with a misspecified base partition, $L_2$ regularization slightly outperforms $L_1$ regularization (see Table~\ref{tab:extVaryRange} where the range parameter varies and Table~\ref{tab:triangle} where the true partition is $\truePtrg$). Although $L_1$ regularization shows higher performance in Table~\ref{tab:extVarySill} when $\sigma^2$ varies for Model~M with a misspecified base partition, the varying $\sigma^2$ values under $\truePsqr$ are more distinct compared to the varying $\phi$ and $\truePtrg$ cases. These results may suggest that $L_2$ regularization is a better choice when fitting models to real data, as it is likely that the base partition will be misspecified under a more complex latent true partition.}




\subsection{True Partition Recovery} \label{TruePartitionRecovery}
To evaluate the accuracy of an estimated partition, we use the Rand index (RI). As a measure of the similarity between two data clusters, the Rand index is defined as the proportion of points that are correctly grouped together based on the true model partition $\trueP$ and the estimated partition $\hat{\mathcal{P}}$, i.e., $\operatorname{RI} = (O_{ss} + O_{dd})/\binom{D}{2}$, where $O_{ss}$ ($O_{dd}$) is the number of pairs of sites that are simultaneously in the same (different) subset of both $\trueP$ and $\hat{\mathcal{P}}$.
We also introduce a local Rand index to assess the partition accuracy for each site. For each specific site $\bm{s}_j$, for $j = 1,\ldots,D$, the $\operatorname{LRI}(\bm{s}_j)$ is defined as
\begin{equation}
\begin{aligned}
    \operatorname{LRI}(\bm{s}_j) = \frac{1}{D-1}\sum_{i\neq j} \mathbb{I}( &\bm{s}_i, \bm{s}_j\in \text{ the same subregion in both } \trueP \text{ and } \hat{\mathcal{P}} \text{ or } \\ 
    &  \bm{s}_i, \bm{s}_j\in \text{ different subregions in both } \trueP \text{ and } \hat{\mathcal{P}}).
\end{aligned}
\end{equation}
where $\mathbb{I}(\cdot)$ denotes the indicator function. Larger values of $\operatorname{LRI}(\bm{s}_j)$ correspond to better classification of site $\bm{s}_j$.

\begin{table}[t!]
    \centering
    \caption{Rand index values for the well-specified and misspecified variants of the true partition designs in Figure~\ref{pic:4truePar}, with both choices of regularization and dependence regimes. The best value in each row and for each setting (well-specified or misspecified) is in bold.}
    \begin{tabular}{l|l|l|l|l|l|l|l}
    \hline
    \multicolumn{2}{c|}{Base partition} & \multicolumn{3}{c|}{Well-specified} & \multicolumn{3}{c}{Misspecified} \\
    \hline
    \multicolumn{2}{c|}{} & Model~B & \multicolumn{2}{c|}{Model~M} & Model~B & \multicolumn{2}{c}{Model~M} \\
    \cline{4-5} \cline{7-8} 
    \multicolumn{2}{c|}{} &  & $L_1$ & $L_2$ &  & $L_1$ & $L_2$ \\
    \hline
    true partition $\truePsqr$ & Case C1  & $0.7598$ & $\bm{0.9880}$ & $0.9879$ & $0.7588$ & $\bm{0.8610}$ & $0.8607$ \\
    (Square) & Case C2 & $0.7598$ & $\bm{0.9835}$ & $0.9786$ & $0.7588$ & $0.8415$ & $\bm{0.8441}$ \\
    \hline
    true partition $\truePtrg$ & Case C1 & - & - & - & $0.7581$ & $0.8031$ & $\bm{0.8051}$ \\
    (Triangular) & Case C2 & - & - & - & $0.7581$ & $0.7706$ & $\bm{0.7748}$ \\
    \hline
    \end{tabular}
    \label{tab:RI}
\end{table}

Table \ref{tab:RI} summarizes the performance of models in terms of their recovery of true partitions $\truePsqr$ and $\truePtrg$. The results show that Model~M generally outperforms Model~B by providing larger Rand index values overall, {with the $L_1$ and $L_2$ regularizations performing almost equivalently under well-specified and misspecified base partitions. Thus, overall, our results in Table \ref{tab:RI} show that Model~M provides major improvements compared to Model~B in terms of partition recovery, but that the specific choice of regularization is less important.}

\begin{figure}[t!]
    \centering
    \includegraphics[width=0.75\textwidth]{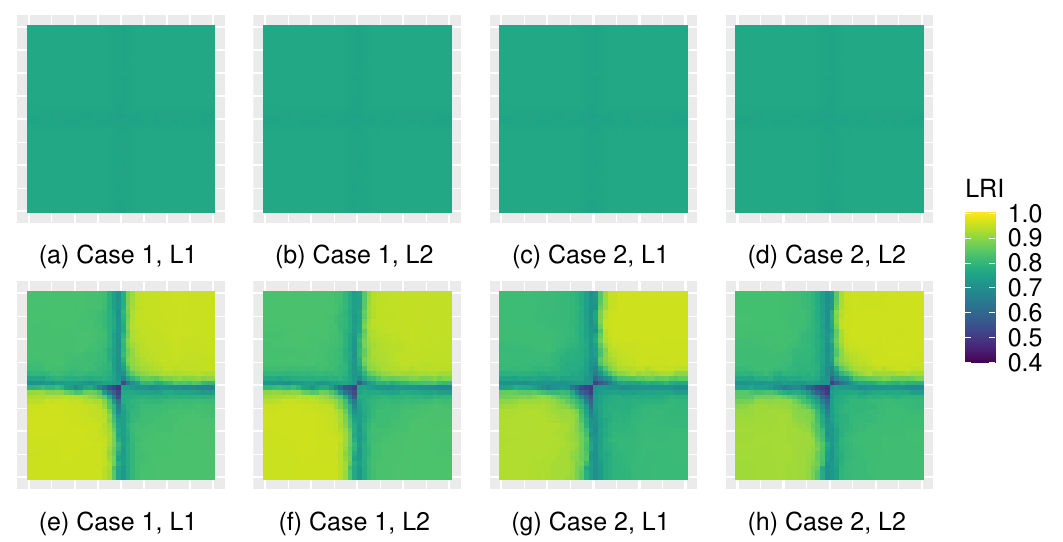} 
    \caption{Average local Rand index for Model~B (first row) and Model~M (second row) for each site with a true partition $\truePsqr$ and misspecified base partition $\mathcal{P}^{(0)}$ (based on the convex hull and $k$-means clustering).}
    \label{pic:LRI+}
\end{figure}


Figures \ref{pic:LRI+}, \ref{pic:LRI+well} and \ref{pic:LRIx} (see the Supplementary Material for the latter two) give the local Rand index (LRI) for each site for the three scenarios discussed in Section \ref{ModelPerformance}, which illustrate the performance of the merged or base partitions on each pixel in terms of region classification. To investigate the sensitivity of Algorithm~\ref{alg:comb} to the choice of base partition, in terms of its efficacy in partition recovery, the simulation study permits a variety of base partitions in a misspecified setting. Model~M with a well-specified base partition generally gives the correct classification of each site, as shown in Figure \ref{pic:LRI+well}. When the base partition is misspecified, the merged partitions generally provide smaller LRI values for the few sites near the boundaries of the true partition and considerably larger LRI values for the many sites within the interior of the true subdomains. This pattern indicates poor classification of sites around the boundaries of the true partition, which may be because these sites are within a subregion of the base partition that overlaps the true boundaries, and the algorithm just merges some subregions in the base partition. In the base partition, the smaller the subregions are, the fewer the sites in the subregions lying on the boundaries, and the merged partition can be less sensitive to the base partition. We note that the LRI values of the sites in subdomains $\mathcal{D}_2$ and $\mathcal{D}_3$ in Figure \ref{pic:4truePar} are smaller than those in subdomains $\mathcal{D}_1$ and $\mathcal{D}_4$ for both true partition cases, as the parameter values in the former subdomains are here assumed to be equal and, hence, they are sometimes merged by our algorithm; this affects partition estimation using Model~M. However, with accurate parameter estimation, such misspecification does not impact our characterization of the extremal dependence structure, and grouping more sites with similar dependence characteristics together in the same cluster may even improve parameter estimation.



\section{Nepal Temperature Data Analysis} \label{Application}

\begin{figure}[t!]
    \centering
    \subfloat{{\includegraphics[width=0.45\textwidth]{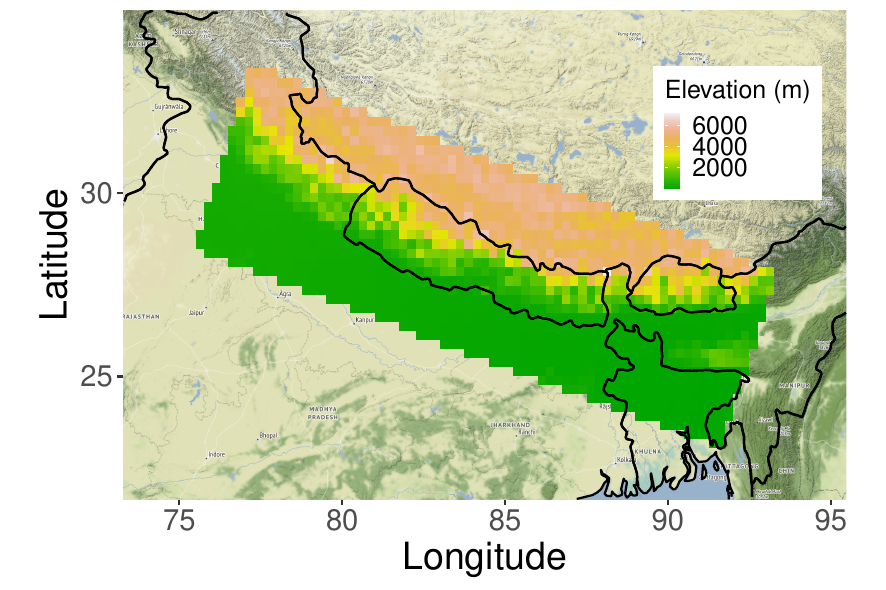} }}%
    \subfloat{{\includegraphics[width=0.45\textwidth]{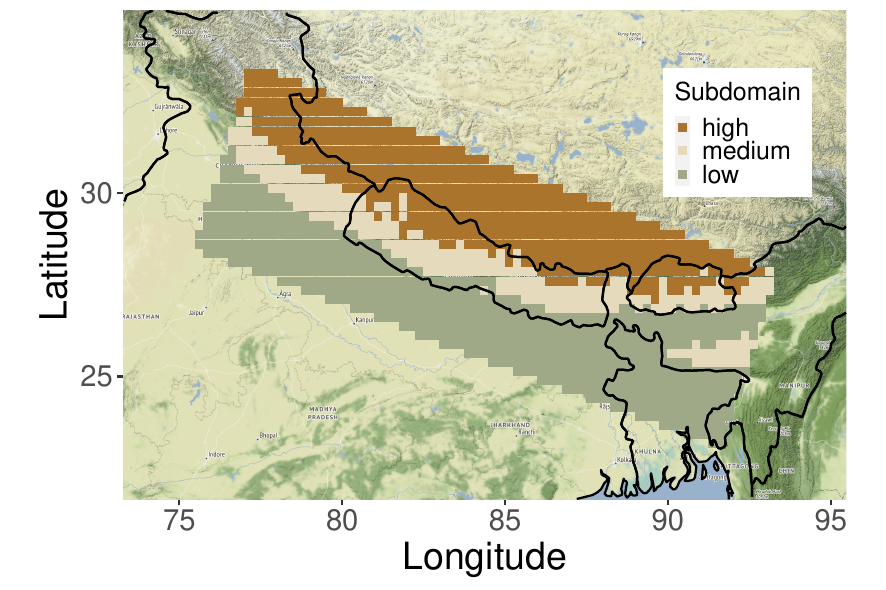} }}%
    \caption{The topographic map of the study region (left) and elevation subdomains (right).}
    \label{pic:Nepal_elev+clu}
\end{figure}

Nepal is a South Asian country covering the Himalayan foothills and the highest peaks of the Himalayan mountains. A strong nonstationary behavior in the dependence of temperature extremes is natural, and to investigate this hypothesis, we obtain daily temperature data, generated by Version 2 of the NASA Global Land Data Assimilation System \citep{rodell2004global}, from 2004 to 2019. We include $D = 1419$ observation sites on a regular grid within and surrounding Nepal (Figure \ref{pic:Nepal_elev+clu}). At these sites, we derive $T = 192$ observations of monthly maxima. 

We use a two-stage procedure for modeling: we first model the marginal behavior of the data before transforming it onto unit-Fréchet margins; we subsequently model the extremal dependence structure using these standardized data. To account for spatial nonstationarity in the margins, we require estimates of site-dependent GEV parameters as in (\ref{eq:GEV}). To accommodate temporal nonstationarity, e.g., seasonality, in the time series, harmonic terms are imposed in the modeling of the $\mu(\bm{s})$ and scale $\varsigma(\bm{s})$. The shape parameter $\xi(\bm{s})$ is fixed with respect to time $t$ for each site. The marginal model for monthly maxima $Y_t(\bm{s}_i)$ at site $\bm{s}_i \in \mathcal{S}$ and month $t$ is $Y_t(\bm{s}_i) \overset{\mathrm{ind}}{\sim} \operatorname{GEV}\{\mu_t(\bm{s}_i), \varsigma_t(\bm{s}_i), \xi(\bm{s}_i)\}$, for all $i = 1,\ldots,D$ and $t= 1,\ldots,T$, with
\begin{equation} \label{spacetime_mu_sigma}
\begin{aligned}
&\mu_t(\bm{s}_i) = \mu_0(\bm{s}_i) + \mu_1(\bm{s}_i)\sin(2\pi t_{month}/12) + \mu_2(\bm{s}_i)\cos(2\pi t_{month}/12), \text{ and} \\
&\log[\varsigma_t(\bm{s}_i)] = \varsigma_0(\bm{s}_i) + \varsigma_1(\bm{s}_i)\sin(2\pi t_{month}/12) + \varsigma_2(\bm{s}_i)\cos(2\pi t_{month}/12),
\end{aligned}
\end{equation} 
where $t_{month}$ is the index of the corresponding month of each observation and $\mu_0(\bm{s})$, $\mu_1(\bm{s})$, and $\mu_2(\bm{s})$ ($\varsigma_0(\bm{s})$, $\varsigma_1(\bm{s})$, and $\varsigma_2(\bm{s})$) are coefficients for covariates of the location (scale) structure. Bayesian inference for spatially-varying marginal parameters may be performed flexibly and quickly using the Max-and-Smooth method, introduced by \cite{hrafnkelsson2021max}, \cite{johannesson2022approximate}, and \cite{hazra2021latent}, where unknown GEV parameters are transformed using specific link functions and then assumed to follow a multivariate Gaussian distribution with fixed and random effects at a latent level. We perform approximate Bayesian inference using the stochastic partial differential equations approach and MCMC methods (see references for full details). Using the posterior mean of each marginal parameter, we transform the original data onto the unit Fr\'echet scale. We draw the quantile-quantile (QQ) plot for transformed data and conduct a hypothesis testing based on Kolmogorov--Smirnov distance, where the results show that the marginal fit is satisfactory overall (more details are in Section~\ref{Supp:MASDiag} in the Supplementary Material). 

\begin{figure}[t!]
    \centering
    \includegraphics[width=0.99\linewidth]{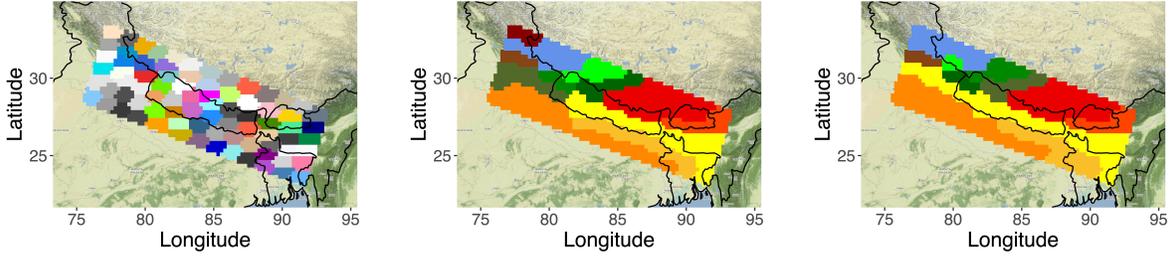}
    \caption{The base partition consisting of 80 subregions (left) and the estimated merged partitions with $L_1$ (middle) and $L_2$ (right) regularization using Brown--Resnick processes, consisting of 12 and 11 subregions, respectively.}
    \label{pic:nepal-BaseAndMerged}
\end{figure}

We fit five Brown--Resnick max-stable models to the standardized data with unit-Fréchet margins: a stationary model (Model S), and our proposed nonstationary model with a base and fully merged partition (Models~B and M), using both $L_1$ and $L_2$ regularization; we compare the performance of all five models. The base partition (Figure \ref{pic:nepal-BaseAndMerged}) consists of $\numOfReg^{(0)}=80$ subregions and we construct them as the convex hull of the clusters determined by applying $k$-means clustering to normalized longitude and latitude coordinates. A validation set ($15\%$) is created to assess model performance by extracting observations from a random sample of all observation sites; the remaining data are used for training. To improve the numerical stability of our optimization algorithm, we exploit ideas of cross-validation and, rather than using the penalized pairwise log-likelihood evaluated on a single holdout set (as described in Section~\ref{DomainPartition}), we split the training data into five folds, {and choose one fold as the holdout set 
so that we fit five models in parallel}; we then tune the smoothness penalty $\bm{\lambda}$ and perform subregion merging by comparing the average pairwise log-likelihood, or penalized pairwise log-likelihood, evaluated over all folds. Cross-validation type is crucial for this small dataset, as the small holdout set sampling may lead to unexpected bias in parameter estimates. The initial descending grid for the $\bm{\lambda}$-tuning procedure is chosen to be the same as in the simulation study. We compute the pairwise log-likelihood on the validation set, the penalized pairwise log-likelihood averaged over folds, and the CLIC and CBIC metrics for the remaining data excluding the validation set. 



\begin{table}[t!]
    \centering
    \caption{Model performance metrics ($\times10^3$) using MSP. The number of parameters for each model is provided in brackets. The best value in each row is in bold.}
    \begin{tabular}{l|c|c|c|c|c}
    \hline
    \multirow{2}{*}{} & \multirow{2}{*}{Model S (2)} & \multicolumn{2}{c|}{$L_1$} & \multicolumn{2}{c}{$L_2$} \\
    \cline{3-6}
    &  & Model B (160) & Model M (24) & Model B (160) & Model M (22) \\
    \hline
    PL & $-18147$ & $-18065$ & $-18044$ & $-18065$ & $\bm{-18039}$ \\
    PPL & -- & $-23325$ & $-23296$ & $-23324$ & $\bm{-23294}$ \\
    CLIC & $1176934$ & $1170592$ & $1169343$ & $1170542$ & $\bm{1168585}$ \\
    CBIC & $1177379$ & $1172061$ & $1171492$ & $1171952$ & $\bm{1169465}$ \\
    \hline
    \end{tabular} 
    \label{tab:Nepal_performance}
\end{table}

\begin{figure}[t!]
    \centering
    \subfloat{{\includegraphics[width=\textwidth]{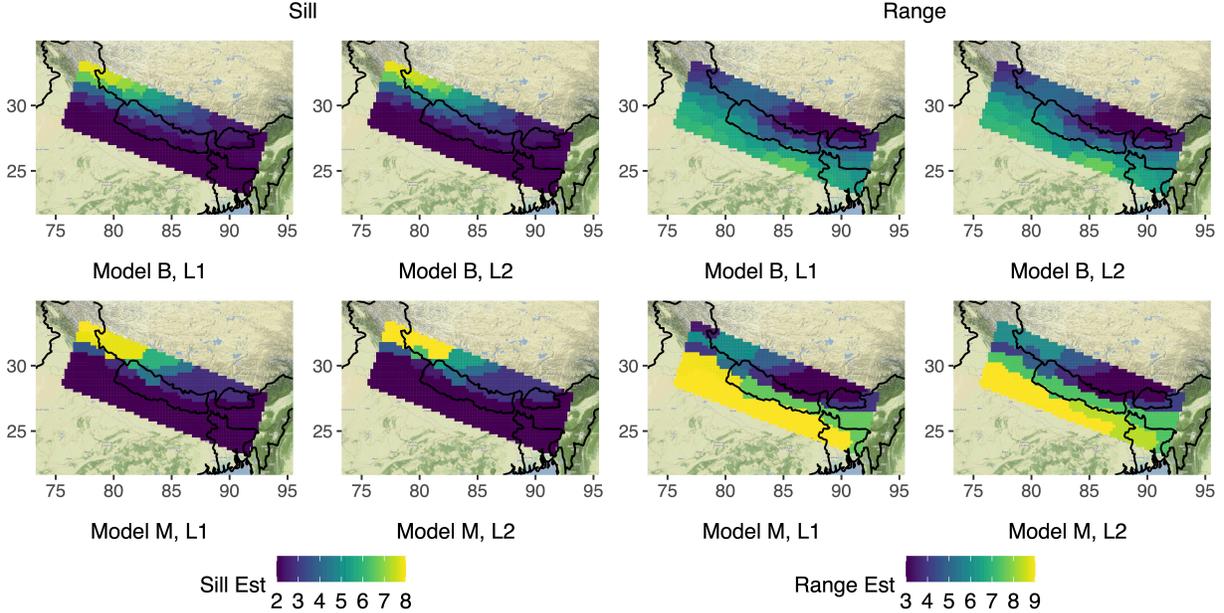} }}
    \caption{Estimated sill (left block) and range (right block) parameters using MSP. Each 2-by-2 block displays Models~B (top) and M (bottom) with $L_1$ (left column) and $L_2$ (right column) regularization.}
    \label{pic:nepal-estimates}
\end{figure}

Table \ref{tab:Nepal_performance} shows that all nonstationary models provide a better fit than the stationary one, as we observe improvement in all goodness-of-fit metrics; we further observe that Model~M outperforms Model~B, regardless of the regularization used. For both Model~B and Model~M, we find that $L_2$ regularization performs slightly better than $L_1$ {in our real data application, which validates the findings from Section~\ref{TruePartitionRecovery}.} Figure \ref{pic:nepal-BaseAndMerged} shows the final estimated partitions for both regularization schemes, while estimates of the sill and range parameters are given in Figure \ref{pic:nepal-estimates}. The partitions roughly identify the plain and mountainous areas shown in Figure \ref{pic:Nepal_elev+clu}, suggesting significantly different extremal dependence behavior in these areas. All nonstationary models provide larger sill estimates and smaller range estimates in mountainous regions than the plains, suggesting that extremal dependence in these areas decays faster with distance; the converse holds for the plains. 



\begin{figure}[hbt!]
    \centering
    \includegraphics[width=14.7cm]{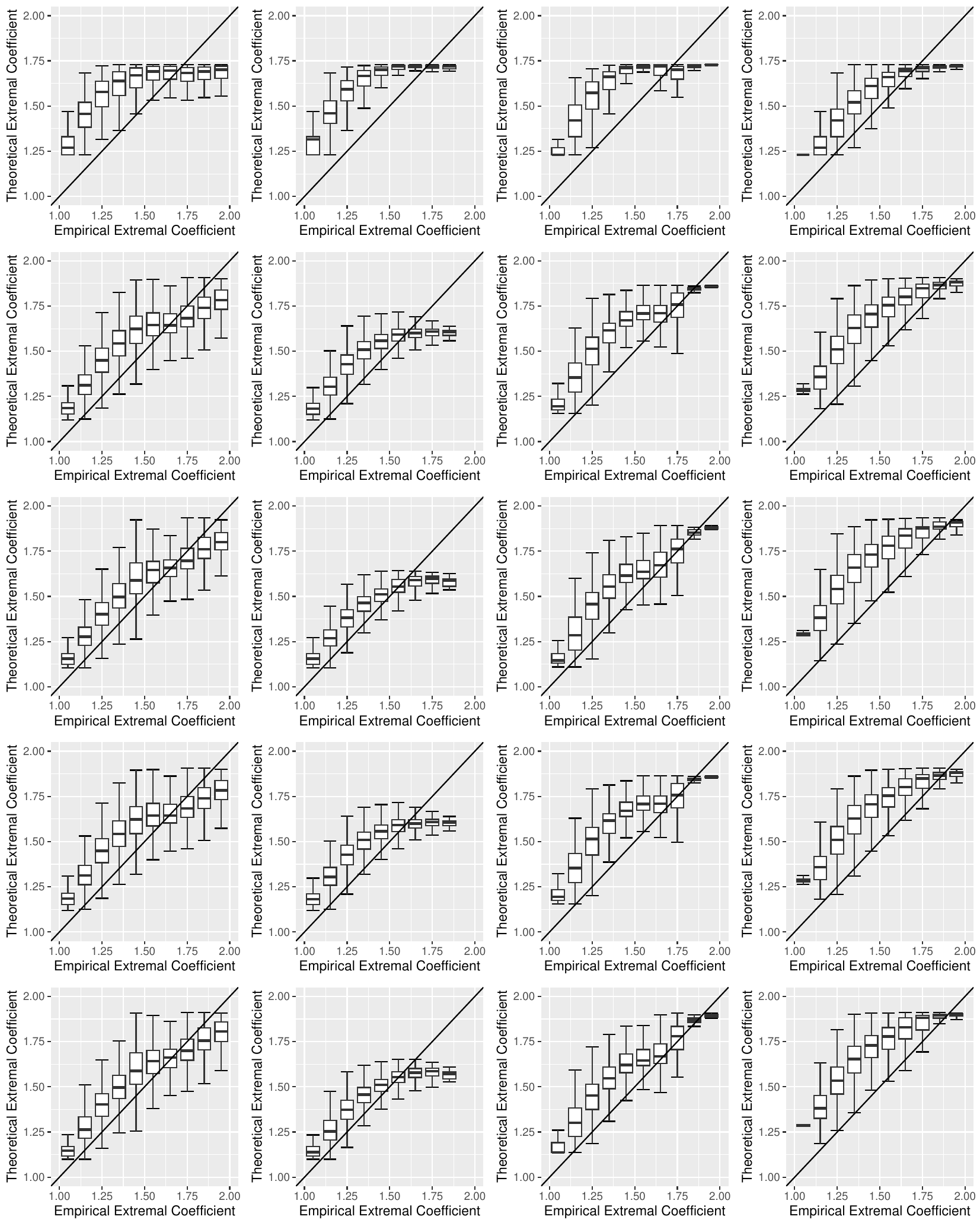}
    \caption{Theoretical extremal coefficients obtained for the stationary and nonstationary Brown--Resnick ($L_1$ and $L_2$ penalty) models (MSP) against empirical extremal coefficients, stratified by elevation subdomain. First column: total; second column: low; third column: medium; fourth column: high. First row: Model~S; second/third row: Model~B with $L_1$/$L_2$ regularization; fourth/fifth row: Model~M with $L_1$/$L_2$ regularization.}
    \label{pic:nepal_extDep_nonsta}
\end{figure}

As a further model diagnostic, we compare the theoretical extremal coefficients, defined in (\ref{eq:extCoef}), from the fitted models with their empirical counterparts; empirical estimates $\widetilde{\nu}(\bm{s}_i, \bm{s}_j)$ are obtained using the $F$-madogram (see Section \ref{Construction}). As the resulting empirical extremal coefficient $\widetilde{\theta}(\bm{s}_i, \bm{s}_j)$ does not necessarily lie in the interval $[1,2]$, we truncate estimates outside of this interval. Computation was conducted using the $\texttt{R}$ package \texttt{SpatialExtremes} \citep{ribatet2011spatialextremes}. To assess model fits, we partition the domain into three subdomains according to low, medium, and high elevation (see Figure \ref{pic:Nepal_elev+clu}), and compute pairwise extremal coefficients for all pairs of sites within each subdomain. Figure~\ref{pic:nepal_extDep_nonsta} displays the theoretical extremal coefficients for the different model fits against their empirical counterparts. We further compute the mean absolute difference between the theoretical and empirical extremal coefficients for all pairs of sites in the whole domain and in each elevation subdomain, and report the results in Table~\ref{tab:Nepal_extDep}.

From Figure \ref{pic:nepal_extDep_nonsta}, we observe that the stationary model fails completely at identifying the varied extremal dependence behavior within each subdomain, while the nonstationary models, regardless of the partition and penalty used, provide a much better description of different extremal dependence behavior. While the theoretical extremal coefficients for the nonstationary models may appear to slightly underestimate the extremal dependence (compared to the empirical ones), it is important to note that this apparent lack of fit may be due to the high variability of the empirical estimates. We here have 192 independent monthly maxima replicates, and hence the empirical extremal coefficients may not be well estimated, especially in weak-dependence settings when $\theta(\bm{s}_i, \bm{s}_j)\approx2$. Another reason might be that the exponential variogram we used has limited flexibility for capturing long-range independence. Model~M better describes nonstationary extremal dependence behavior overall, especially in the low-elevation subdomain, which is confirmed by the mean absolute difference (MAD) values; see Table \ref{tab:Nepal_extDep}. Nonstationary models significantly outperform the stationary model in the low-elevation and middle-elevation subdomains in terms of MAD, with Model~M slightly outperforming Model~B. In the middle-elevation subdomain, models S and B are comparable, but Model~M provides the smallest MAD. In the high-elevation subdomain, Model~S {provides smaller MAD values than the nonstationary models, which may be because nonstationary models provide more realistic (larger) fitted values with increasing uncertainty in the high-elevation subdomain.} However, Model~M again outperforms Model~B. Finally, the total MAD between theoretical and empirical extremal coefficients for the whole domain is smaller for the nonstationary models, with Model~M outperforming Model~B. Therefore, we conclude that the nonstationary models provide better estimates of the extremal dependence structure of air temperature within Nepal and its surrounding Himalayan and sub-Himalayan regions, and our proposed merging procedure can further improve the fit, with $L_1$ regularization generally outperforming $L_2$ regularization.


\begin{table}[t!]
    \centering
    \caption{Mean absolute differences for pairwise extremal coefficients using MSP, with pairs in the whole domain (``Total'') or stratified by elevation subdomain (``Low'', ``Medium'', and ``High'').}
    \begin{tabular}{l|c|c|c|c|c}
    \hline
    \multirow{2}{*}{} & \multirow{2}{*}{Model S} & \multicolumn{2}{c|}{$L_1$} & \multicolumn{2}{c}{$L_2$} \\
    \cline{3-6}
    &  & Model~B & Model~M & Model~B & Model~M \\
    \hline
    Total & 0.1701 & 0.1322 & \textbf{0.1183} & 0.1321 & 0.1193 \\
    Low & 0.2406 & 0.121 & 0.0915 & 0.1209 & \textbf{0.0885} \\
    Medium & 0.2393 & 0.2139 & \textbf{0.1702} & 0.2136 & 0.1717 \\
    High & \textbf{0.1154} & 0.1854 & 0.2085 & 0.1857 & 0.2061 \\
    \hline
    \end{tabular} 
    \label{tab:Nepal_extDep}
\end{table}


To assess the sensitivity of the estimated partition to the choice of extremal dependence model, we replace the Brown--Resnick process with an inverted Brown--Resnick process \citep{wadsworth2012dependence}; the details for this process and its inference are provided in Section~\ref{Supp:IMSP} of the Supplementary Material. We find that the inverted Brown--Resnick model fits result in a similar partitioning of the spatial domain to that shown in Figure~\ref{pic:nepal-estimates}. {However, choosing an inverted max-stable model over a max-stable one provides a less preferable fit overall, with smaller PL, PPL, CLIC, and CBIC for the merged models. Hence, we omit full details of these results for brevity.}

\section{Concluding Remarks} \label{ConcludingRemarks}

In this paper, we propose a flexible yet parsimonious model for nonstationary extremal dependence using the Brown--Resnick max-stable process. We develop a new computationally-efficient method to simultaneously estimate unknown parameters, identify stationary subregions, and reduce the number of model parameters. More precisely, our model uses a nonstationary variogram and a partitioning of the study domain into locally-stationary subregions. We further propose practical strategies to merge subregions according to parameter estimates, which create a parsimonious model in a data-driven way without expert knowledge. We draw inferences using a pairwise likelihood to mitigate the computational expense and investigate statistical efficiency. Dependence regularization that smooths the parameter estimates is also incorporated in our subregion merging algorithm, and the smoothing parameters $\bm{\eleOfDG}$ are tuned adaptively. In both our simulation study and application, we show that our model provides good fits to data and better captures nonstationarity in extremal dependence, significantly improving over models in the existing literature.

Our simulation study shows that, even when the base partition is misspecified, we still observe improvement in model fit over stationary models, but even better fits are observable with a well-specified base partition. Choosing the base partition for our model can be problematic, as subregions in the base partition may overlap with the boundaries in the true partition (if one exists) and it can also bring outlier sites with different extremal dependence behavior into the same subregion. Although the algorithm only merges subregions in the base partition, our assessment of the partition recovery in the simulation study (with the local rand index; Section~\ref{TruePartitionRecovery}) suggests that the final partition is not overly sensitive to the initial base partition. For further mitigation, we could divide the domain into finer subregions; however, this may lead to unreliable parameter estimation. Another possible approach could be to allow subregions to split at each step and merge both. Another problem emerges when we create a base partition consisting of subregions that include sites with distinct geographical characteristics compared to the neighbors, e.g., the high peaks of the Himalayas in the neighborhood of the valley areas of Nepal. Here, their extremal dependence structure is dissimilar to the nearby sites. Algorithm~\ref{alg:comb} lacks the capability of identifying this issue. Including covariates (like elevation) in the model may be helpful. However, this would increase the number of model parameters, which contradicts our target of a parsimonious model.

{Our nonstationary Brown--Resnick max-stable process model uses an exponential variogram, which is bounded above as the distance between sites $\bm{s}_i$ and $\bm{s}_j$ increases. Hence, our model can only capture asymptotic dependence between sites; it cannot capture situations where the process observed at two sufficiently distant sites exhibits asymptotic independence. In a relatively small domain of study, asymptotic dependence everywhere may be a reasonable assumption to make, but this may not be the case for large regions; in this case, our model may only be appropriate for smaller subdomains}. Creating valid unbounded nonstationary variograms would make our model more flexible, but this would not allow us to capture forms of asymptotic independence other than perfect independence; a similar problem is discussed in \cite{chevalier2021modeling}. However, asymptotic independence can be accommodated by using inverted MSPs \citep{wadsworth2012dependence} (as in  Section~\ref{Supp:IMSP} of the Supplementary Material) or adapting our methodology to other types of asymptotic independence models, such as certain types of Gaussian location-scale mixtures \citep{opitz2016modeling, huser2017bridging, hazra2022realistic, zhang2022modeling}, which can capture sub-asymptotic extremal dependence \citep{huser2022advances}. 

Further future work includes an extension to incorporate local anisotropy and adapting our methodology to other max-stable processes, e.g., extremal-$t$; the former can be easily accommodated by parameterizing the matrix $\Omega(\bm{s})$ in (\ref{nonsta-rho}) and the latter via replacement of the likelihood function in \eqref{pairwiseLike}. Our approach can also be extended to other spatial extremes dependence models, such as $r$-Pareto processes \citep{de2018high} or the conditional model of \citet{WADSWORTH2022100677}. While our methodology is frequentist, rather than Bayesian, despite the differences in the paradigms one may consider the inclusion of the penalty term in the likelihood as a prior choice. Moreover, the solution obtained by maximizing the underlying objective function coincides with the maximum a-posterior estimate \citep{casella2010penalized}. Thus, a fully Bayesian formulation of our methodology is also possible.

{For reproducibility purposes, the Supplementary Material contains our \texttt{R} code, the gridded Nepal temperature data, and a small simulation example, which can also be obtained from the following Github link: \url{https://github.com/Xuanjie-Shao/NonStaExtDep}.}

\section*{Acknowledgement}
The authors would like to thank an Editor, an Associate Editor, and two anonymous reviewers for their several thoughtful suggestions which substantially improved the paper. This publication is based upon work supported by the King Abdullah University of Science and Technology (KAUST) Office of Sponsored Research (OSR) under Award No. OSR-CRG2020-4394. The research of the second author is partially supported by the Indian Institute of Technology Kanpur and Rice University collaborative research grant under Award No. DOIR/2023246.

\bigskip
\begin{center}
{\large\bf Supplementary Material}
\end{center}

\begin{description}

\item[PDF Supplement:] This supplement contains further results obtained in our simulation studies (optimization time and accuracy) and the data application (marginal goodness-of-fit diagnostics and some results based on fitting inverted max-stable processes) (PDF file)

\item[\texttt{R} Code:] This supplement contains the codes for implementing our proposed method in \texttt{R}, the gridded temperature data (both daily average and monthly maxima) of Nepal and its surrounding regions analyzed in this paper, and a small simulation example (zip file)

\end{description}


\spacingset{1}

\bibliographystyle{Chicago}
\bibliography{Bibliography}

\end{document}